\documentclass[a4paper, 10 pt, conference]{ieeeconf}
\IEEEoverridecommandlockouts
\usepackage{amsfonts,amsmath,amssymb} % assumes amsmath package installed

\usepackage{psfrag,color}
\usepackage{enumerate,cite,latexsym,graphicx}
\newtheorem{theorem}{Theorem}
\newtheorem{lemma}{Lemma}

\def\tr{\mathop{\rm Tr}\nolimits} % Trace of matrix

\title{\LARGE \bf Guaranteed Non-quadratic Performance for Quantum Systems with Nonlinear Uncertainties
}

\author{Ian R.~Petersen %
\thanks{This work was supported by the
Australian Research Council (ARC) and the Air Force Office of Scientific
Research (AFOSR). This material is based on research sponsored by the
Air Force Research Laboratory, under agreement number FA2386-12-1-4075.  The U.S. Government is authorized to reproduce and
distribute reprints for Governmental purposes notwithstanding any
copyright notation thereon.
The views and conclusions contained herein are those of the authors
and should not be interpreted as necessarily representing the official
policies or endorsements, either expressed or implied, of the Air
Force Research Laboratory or the U.S. Government. }%
\thanks{Ian R. Petersen is with the School of  Engineering and Information Technology, 
        University of New South Wales at the Australian Defence Force Academy, Canberra ACT 2600, Australia.
         {\tt\small i.r.petersen@gmail.com} } 
}%

\begin{document}

\maketitle
\thispagestyle{empty}
\pagestyle{empty}

\begin{abstract}
This paper presents a robust performance analysis result for a class of uncertain quantum systems containing sector bounded nonlinearities arising from perturbations to the system Hamiltonian. An LMI condition is given for calculating a guaranteed upper bound on a non-quadratic cost function. This result is illustrated with an example involving a Josephson junction in an electromagnetic cavity. 
\end{abstract}

%%%%%%%%%%%%%%%%%%%%%%%%%%%%%%%%%%%%%%%%%%%%%%%%%%%%%%%%%%%%%%%%%%%%%%%%%%%%%%%%
\section{Introduction} \label{sec:intro}
A number of papers have considered in recent years, the feedback
control of systems  governed by the laws of quantum
mechanics rather than systems  governed by the laws of classical mechanics; e.g., see
\cite{YK03A,YK03B,YAM06,JNP1,NJP1,GGY08,MaP3,MaP4,YNJP1,GJ09,GJN10,WM10,BR04,PET10Ba,ShP5,PET08A,PET09A,PET10A,PET10Ca}. In
particular, the papers \cite{GJ09,JG10} consider a framework of
quantum systems defined in terms of a triple $(S,L,H)$ where $S$ is a
scattering matrix of operators, $L$ is a vector of coupling operators and $H$ is a
Hamiltonian operator. All operators are on an underlying Hilbert space. 

The paper \cite{PUJ1a} considers a quantum system defined by a triple $(S,L,H)$ such that the quantum system Hamiltonian is
written as $H =H_1+H_2$. Here $H_1$ is a known nominal Hamiltonian
and $H_2$ is a perturbation Hamiltonian, which is contained in a
set of Hamiltonians $\mathcal{W}$. The paper \cite{PUJ1a} considers a problem of absolute stability for such uncertain quantum systems for the case in which the
nominal Hamiltonian $H_1$ is a quadratic function of annihilation and
creation operators and the coupling operator vector $L$ is a linear
function of annihilation and creation operators. Such as nominal quantum system is said to be a  linear quantum system; e.g., see
\cite{JNP1,NJP1,MaP3,MaP4,PET10Ba}. However, the perturbation Hamiltonian $H_2$ is assumed to be contained in a set of non-quadratic  Hamiltonians corresponding to a sector bounded nonlinearity. Then, the paper \cite{PUJ1a} obtains a frequency domain robust stability result. Extensions of the approach of \cite{PUJ1a} can be found in the papers \cite{PUJ2,PUJ3a,JPU1a,PET12Aa,PET12Ba,PET13Aa,PET13Ba} in which similar robust stability results are of obtain for uncertain quantum systems with different classes of uncertainty and different applications to specific quantum systems. Also, in the paper \cite{JPU1a} a problem of robust performance analysis as well as robust stability analysis is considered. 

In this paper, we extend the results of \cite{PUJ1a,JPU1a,PET12Aa} by considering a problem of robust performance analysis with a non-quadratic cost functional for the class of uncertain quantum systems of the form considered in \cite{PUJ1a,PET12Aa}. The motivation for considering robust performance of a quantum system with a non-quadratic cost function arises from the fact that the presence of nonlinearities in the quantum system allows for the possibility of a non-Gaussian system state; e.g., see \cite{GZ00}. Such non-Gaussian system states include important non-classical states such as the Schr\"odinger cat state (also known as a superposition state, e.g., see \cite{WM08}). These non-classical quantum states are useful in areas such as quantum information and quantum communications; e.g., see \cite{NC00}. The presence of such non-classical states can be verified by obtaining a suitable bound on a non-quadratic cost function (such as the Wigner function, e.g., see \cite{GZ00,WM08}). Our approach to obtaining a bound on the non-quadratic cost function is to extend the sector bound method considered in \cite{PUJ1a} to bound both the nonlinearity and non-quadratic cost function together. It is important that these two quantities are bounded together since the non-Gaussian state only arises due to the presence of the nonlinearity in the quantum system dynamics. Then, by applying a similar approach to that in \cite{PUJ1a,JPU1a} we are able to derive a guaranteed upper bound on the non-quadratic cost function in terms of an LMI problem. In order to illustrate this result, it is applied to an example of a quantum system consisting of a Josephson junction in an electromagnetic cavity. The robust stability of a similar  system was previously considered in the paper \cite{PET12Aa}. In this paper, we consider the robust performance of this system with respect to a non-quadratic cost functional. 

A future application of the robust performance analysis approach proposed in this paper would be to use it to develop a method for  the design of coherent quantum feedback controllers for quantum systems to achieve a certain  closed loop performance bound in terms of a non-quadratic cost functional. In such a coherent quantum feedback control scheme both the plant and controller are quantum systems; e.g., see \cite{NJP1}. This would be useful in the generation of non-classical quantum states which are needed in areas of quantum computing and quantum information; e.g., see \cite{NC00}.

\section{Quantum Systems with Nonlinear  Uncertainties} \label{sec:systems}
The parameters $(S,L,H)$ will be considered to define an uncertain nonlinear  quantum system. Here,  $S$ is the scattering matrix, which is  chosen as the identity matrix, L is the coupling operator vector and $H$ is the system  Hamiltonian operator. $H$ is assumed to be of the form
\begin{equation}
\label{H1}
H = \frac{1}{2}\left[\begin{array}{cc}a^\dagger &
      a^T\end{array}\right]M
\left[\begin{array}{c}a \\ a^\#\end{array}\right]+f(z,z^*).
\end{equation}
Here, $a$ is an $n$-dimensional vector of annihilation
operators on the underlying Hilbert space and $a^\#$ is the
corresponding vector of creation operators. Also, $M \in \mathbb{C}^{2n\times 2n}$ is a Hermitian matrix of the
form
\begin{equation}
\label{Mform}
M= \left[\begin{array}{cc}M_1 & M_2\\
M_2^\# &     M_1^\#\end{array}\right]
\end{equation}
and $M_1 = M_1^\dagger$, $M_2 = M_2^T$.
In the case of vectors of
operators, the notation  $^\dagger$ refers to the transpose of the vector of adjoint
operators and  in the case of matrices, this notation refers to the complex conjugate transpose of a matrix. In the case of vectors of
operators, the notation $^\#$ refers to the vector of adjoint
operators and in the case of complex matrices, this notation refers to
the complex conjugate matrix. Also, the notation $^*$ denotes the adjoint of an
operator. The matrix $M$ is assumed to be known and defines the nominal quadratic part of the system Hamiltonian. 
  Furthermore, we assume the uncertain non-quadratic  part of the system Hamiltonian  $f(z,z^*)$ is defined by a formal power series of  the form
\begin{eqnarray}
\label{H2nonquad}
f(z,z^*)
&=& \sum_{k=0}^\infty\sum_{\ell=0}^\infty S_{k\ell}z^k(z^*)^\ell\nonumber \\
&=& \sum_{k=0}^\infty\sum_{\ell=0}^\infty S_{k\ell}H_{k\ell},
\end{eqnarray}
which is assumed to converge in some suitable sense.
Here $S_{k\ell}=S_{\ell k}^*$, $H_{k\ell} = z^k(z^*)^\ell$,  and $z$ is a known scalar operator defined by
\begin{eqnarray}
\label{z}
z &=&  E_1a+E_2 a^\# \nonumber \\
&=& \left[\begin{array}{cc} E_1 & E_2 \end{array}\right]
\left[\begin{array}{c}a \\ a^\#\end{array}\right] = \tilde E 
\left[\begin{array}{c}a \\ a^\#\end{array}\right];
\end{eqnarray}
i.e., the vector $\tilde E \in \mathbb{C}^{1\times 2n}$ is a known complex vector.

The term $f(z,z^*)$ is referred to as the perturbation Hamiltonian. It  is assumed to be unknown but is contained within a known set which will be defined below.

We assume the coupling operator vector $L$  is known and is of the form 
\begin{equation}
\label{L}
L = \left[\begin{array}{cc}N_{1} & N_{2}\end{array}\right]\left[\begin{array}{c}a \\ a^\#\end{array}\right].
\end{equation}
Here,  $N_{1} \in \mathbb{C}^{m\times n}$,  $N_{2} \in
\mathbb{C}^{m\times n}$ are known matrices. 
Also, we write
\begin{eqnarray*}
\left[\begin{array}{c}L \\ L^\#\end{array}\right] &=& N
\left[\begin{array}{c}a \\ a^\#\end{array}\right] \\
&=&
\left[\begin{array}{cc}N_{1} & N_{2}\\
N_{2}^\# &     N_{1}^\#\end{array}\right] 
\left[\begin{array}{c}a \\ a^\#\end{array}\right].
\end{eqnarray*} 

The annihilation and creation operators $a$ and $a^\#$ are assumed to satisfy the
canonical commutation relations:
\begin{eqnarray}
\label{CCR2}
\left[\left[\begin{array}{l}
      a\\a^\#\end{array}\right],\left[\begin{array}{l}
      a\\a^\#\end{array}\right]^\dagger\right]
&\stackrel{\Delta}{=}&\left[\begin{array}{l} a\\a^\#\end{array}\right]
\left[\begin{array}{l} a\\a^\#\end{array}\right]^\dagger
\nonumber \\
&&- \left(\left[\begin{array}{l} a\\a^\#\end{array}\right]^\#
\left[\begin{array}{l} a\\a^\#\end{array}\right]^T\right)^T\nonumber \\
&=& J
\end{eqnarray}
where $J = \left[\begin{array}{cc}I & 0\\
0 & -I\end{array}\right]$; e.g., see \cite{GGY08,GJN10,PET10Ba}. 

Also, we will consider a non-quadratic cost defined as
\begin{equation}
\label{W}
\mathcal{C} = \limsup_{T\to\infty}\frac{1}{T}\int_0^T \langle W(z(t),z(t)^*)\rangle dt
\end{equation}
where $W(z,z^*)$ is a suitable non-quadratic function. Here $z(t)$, $z(t)^*$, denotes the Heisenberg evolution of the operators $z$, $z^*$ and
$\langle\cdot\rangle$ denotes quantum expectation; e.g., see \cite{JG10}. The non-quadratic function $W(z,z^*)$ is assumed to satisfy the following quadratic upper bound condition:
\begin{equation}
\label{Wbound}
W(z,z^*)\leq \frac{1}{\gamma_0^2}z z^* + \delta_0,
\end{equation}
where $\gamma_0 > 0$, $\delta_0\geq 0$ are given constants. 
 $W(z,z^*)$ will also be used in the definition of the set of allowable perturbation Hamiltonians $f(\cdot)$.

To define the set of allowable perturbation Hamiltonians $f(\cdot)$, we first define the following formal partial derivatives:
\begin{equation}
\label{fdash}
\frac{\partial f(z,z^*)}{\partial z} \stackrel{\Delta}{=} \sum_{k=1}^\infty\sum_{\ell=0}^\infty k S_{k \ell}z^{k-1}(z^*)^\ell;
\end{equation}
\begin{equation}
\label{fddash}
\frac{\partial^2 f(z,z^*)}{\partial z^2} 
\stackrel{\Delta}{=} \sum_{k=1}^\infty\sum_{\ell=0}^\infty k(k-1)S_{k\ell} z^{k-2}(z^*)^{\ell}.
\end{equation}
and for given constants $\gamma_1 > 0$, $\gamma_2 > 0$, $\delta_1\geq 0$, $\delta_2\geq 0$,  $\delta_3\geq 0$, we consider the 
sector bound conditions
\begin{equation}
\label{sector4a}
W(z,z^*)+\frac{\partial f(z,z^*)}{\partial z}^*\frac{\partial f(z,z^*)}{\partial z}  
\leq \frac{1}{\gamma_1^2}z z^* + \delta_1,
\end{equation}
\begin{equation}
\label{sector4c}
\frac{\partial f(z,z^*)}{\partial z}^*\frac{\partial f(z,z^*)}{\partial z}  
\leq \frac{1}{\gamma_2^2}z z^* + \delta_2
\end{equation}
and the condition
\begin{equation}
\label{sector4b}
\frac{\partial^2 f(z,z^*)}{\partial z^2}^*\frac{\partial^2 f(z,z^*)}{\partial z^2} \leq  \delta_3.
\end{equation}

Then we define the set of perturbation Hamiltonians $\mathcal{W}$  as follows:
\begin{equation}
\label{W5}
\mathcal{W} = \left\{\begin{array}{l}f(\cdot) \mbox{ of the form
      (\ref{H2nonquad}) such that 
} \\
\mbox{ conditions (\ref{sector4a}), (\ref{sector4c}) and (\ref{sector4b}) are satisfied}\end{array}\right\}.
\end{equation}
Note that the condition (\ref{sector4b}) effectively amounts to a global Lipschitz condition on the quantum nonlinearity. 

Our main result, which gives an upper bound on the non-quadratic cost function (\ref{W}), will be given in terms of the following LMI condition dependent on a parameter $\tau_1 > 0$:
\begin{eqnarray}
\label{LMI}
\left[\begin{array}{cc}
F^\dagger P + P F 
+ \kappa \Sigma \tilde E^T \tilde E^\# \Sigma & 2 PJ\Sigma \tilde E^T  \\
2\tilde E^\# \Sigma JP & -\frac{I}{\tau_1^2} 
\end{array}\right] < 0
\end{eqnarray}
where $\Sigma = \left[\begin{array}{cc} 0 & I\\
I &0 \end{array}\right]$, $F = -\imath JM-\frac{1}{2}JN^\dagger J N$ and the quantity $\kappa > 0$ is defined as
\begin{equation}
\label{kappa}
\kappa = \left\{\begin{array}{ll} 
\frac{1}{\gamma_1^2} +\left(\frac{1}{\tau_1^2}-1\right) & \mbox{ for }
\tau_1^2 \leq 1; \\
\frac{1}{\tau_1^2\gamma_1^2} +\frac{1}{\gamma_0^2}\left(1-\frac{1}{\tau_1^2}\right)& \mbox{ for }
\tau_1^2 >  1.
\end{array}\right. \nonumber \\
\end{equation}

\begin{theorem}
\label{T1}
Consider an uncertain open nonlinear quantum system defined by $(S,L,H)$ and a non-quadratic cost function $\mathcal{C}$
 such that
$H$ is of the form (\ref{H1}), $L$ is of the
form (\ref{L}) and $f(\cdot) \in \mathcal{W}$. Also, assume that $\mathcal{C}$ defined in (\ref{W}) is such that (\ref{Wbound}) is satisfied. Furthermore, assume that there exists a constant $\tau_1 > 0$ such that the LMI (\ref{LMI}) has a solution $P > 0$. Then the
cost $\mathcal{C}$ satisfies the bound:
\begin{equation}
\label{Cbound}
\mathcal{C} \leq \tr\left(PJN^\dagger\left[\begin{array}{cc}I & 0 \\ 0 & 0 \end{array}\right]NJ\right) + \zeta 
+\sqrt{\delta_3}|\mu| 
\end{equation}
where
\begin{equation}
\label{zeta}
\zeta = \left\{\begin{array}{ll} 
\delta_1  + \left(\frac{1}{\tau_1^2}-1\right)\delta_2
 & \mbox{ for } \tau_1^2 \leq 1; \\
\frac{1}{\tau_1^2}\delta_1  + \left(1-\frac{1}{\tau_1^2}\right)\delta_0
& \mbox{ for }
\tau_1^2 >  1
\end{array}\right. \nonumber \\
\end{equation}
and 
\begin{equation}
\label{mu0}
\mu = -\tilde E \Sigma JPJ\tilde E^T.
\end{equation}
\end{theorem}

In order to prove this theorem, we require the following  lemmas.

\begin{lemma}[See Lemma 2 of \cite{JPU1a}]
\label{L0}
Consider an open quantum system defined by $(S,L,H)$ and suppose there exists a non-negative self-adjoint operator $V$ on the underlying Hilbert space such that
\begin{equation}
\label{lyap_ineq}
-\imath[V,H] + \frac{1}{2}L^\dagger[V,L]+\frac{1}{2}[L^\dagger,V]L + W(z,z^*) \leq \lambda
\end{equation}
where $c > 0$ and $\lambda$ are real numbers. 
Then for any system state, we have
\[
\limsup_{T\to\infty}\frac{1}{T}\int_0^T \langle W(t)\rangle dt \leq
\lambda. 
\]
\end{lemma}

We will consider quadratic ``Lyapunov'' operators  $V$ of the form 
\begin{equation}
\label{quadV}
V = \left[\begin{array}{cc}a^\dagger &
      a^T\end{array}\right]P
\left[\begin{array}{c}a \\ a^\#\end{array}\right]
\end{equation}
where $P \in \mathbb{C}^{2n\times 2n}$ is a positive-definite Hermitian matrix of the
form
\begin{equation}
\label{Pform}
P= \left[\begin{array}{cc}P_1 & P_2\\
P_2^\# &     P_1^\#\end{array}\right].
\end{equation}
 Hence, we consider a set of  non-negative self-adjoint operators
$\mathcal{P}$ defined as
\begin{equation}
\label{P1}
\mathcal{P} = \left\{\begin{array}{l}V \mbox{ of the form
      (\ref{quadV}) such that $P > 0$ is a 
} \\
\mbox{  Hermitian matrix of the form (\ref{Pform})}\end{array}\right\}.
\end{equation}

\begin{lemma}[See Lemma 5 in \cite{PUJ1a}]
\label{L1}
Given any $V \in \mathcal{P}$, then
\begin{equation}
\label{mu}
\left[z,[z,V]\right] = \left[z^*,[z^*,V]\right]^* = \mu 
\end{equation}
where the constant $\mu $ is defined as in (\ref{mu0}).
\end{lemma}

\begin{lemma}[See Lemma 3 in \cite{PET13Aa} and Lemma 2 in \cite{PET13Ba}]
\label{L2}
Given any $V \in \mathcal{P}$, then
\begin{eqnarray}
\label{comm_condition}
[V,f(z,z^*)] &=&[V,z ]w_{1}^* -w_{1}[z^*,V]\nonumber \\
&&+ \frac{1}{2}\mu  w_{2}^*-\frac{1}{2}w_{2}\mu^*
\end{eqnarray} 
where
\begin{eqnarray}
\label{zw1w2}
w_1&=&  = \frac{\partial f(z,z^*)}{\partial z }^*,\nonumber \\
w_2&=&  = \frac{\partial^2 f(z,z^*)}{\partial z ^2}^*,\nonumber \\
\end{eqnarray}
and the constant $\mu $ is defined as in (\ref{mu0}). 
\end{lemma}

\begin{lemma}[See Lemma 4 in \cite{PET13Aa}]
\label{L3}
Given $V \in \mathcal{P}$ and $L$ defined as in (\ref{L}), then
\begin{eqnarray*}
\lefteqn{[V,\frac{1}{2}\left[\begin{array}{cc}a^\dagger &
      a^T\end{array}\right]M
\left[\begin{array}{c}a \\ a^\#\end{array}\right]] =}\nonumber \\
&& \left[\left[\begin{array}{cc}a^\dagger &
      a^T\end{array}\right]P
\left[\begin{array}{c}a \\ a^\#\end{array}\right],\frac{1}{2}\left[\begin{array}{cc}a^\dagger &
      a^T\end{array}\right]M
\left[\begin{array}{c}a \\ a^\#\end{array}\right]\right] \nonumber \\
&=& \left[\begin{array}{c}a \\ a^\#\end{array}\right]^\dagger 
\left[
PJM - MJP 
\right] \left[\begin{array}{c}a \\ a^\#\end{array}\right].
\end{eqnarray*}
Also,
\begin{eqnarray*}
\lefteqn{\frac{1}{2}L^\dagger[V,L]+\frac{1}{2}[L^\dagger,V]L =} \nonumber \\
&=& \tr\left(PJN^\dagger\left[\begin{array}{cc}I & 0 \\ 0 & 0 \end{array}\right]NJ\right)
\nonumber \\&&
-\frac{1}{2}\left[\begin{array}{c}a \\ a^\#\end{array}\right]^\dagger
\left(N^\dagger J N JP+PJN^\dagger J N\right)
\left[\begin{array}{c}a \\ a^\#\end{array}\right].
\end{eqnarray*}
Furthermore, 
\[
\left[\left[\begin{array}{c}a \\ a^\#\end{array}\right],\left[\begin{array}{cc}a^\dagger &
      a^T\end{array}\right]P
\left[\begin{array}{c}a \\ a^\#\end{array}\right]\right] = 2JP\left[\begin{array}{c}a \\ a^\#\end{array}\right].
\]
\end{lemma}

\noindent
{\em Proof of Theorem \ref{T1}.}
It follows from (\ref{z}) that we can write
\begin{eqnarray*}
z^* &=& E_1^\#a^\#+E_2^\# a=\left[\begin{array}{cc} E_2^\# & E_1^\# \end{array}\right]
\left[\begin{array}{c}a \\ a^\#\end{array}\right]\nonumber \\
&=&  \tilde E^\# \Sigma \left[\begin{array}{c}a \\ a^\#\end{array}\right].
\end{eqnarray*}
Also,  it follows from Lemma \ref{L3} that
\[
[z^*,V] = 2 \tilde E^\# \Sigma
JP\left[\begin{array}{c}a \\ a^\#\end{array}\right].
\]
Furthermore, $[V,z] = [z^*,V]^*$ and  hence,
\begin{eqnarray}
\label{VzzV}
%\lefteqn{
[V,z] [z^*,V] =%}\nonumber \\&& 
4\left[\begin{array}{c}a \\ a^\#\end{array}\right]^\dagger PJ 
\Sigma \tilde E^T \tilde E^\# \Sigma
JP
\left[\begin{array}{c}a \\ a^\#\end{array}\right].
\end{eqnarray}
Also, we can write
\begin{equation}
\label{zz}
zz^* = \left[\begin{array}{c}a \\ a^\#\end{array}\right]^\dagger
\Sigma \tilde E^T \tilde E^\# \Sigma
\left[\begin{array}{c}a \\ a^\#\end{array}\right].
\end{equation}

Hence using Lemma \ref{L3}, we obtain
\begin{eqnarray}
\label{lyap_ineq3}
&&-\imath[V,\frac{1}{2}\left[\begin{array}{cc}a^\dagger &
      a^T\end{array}\right]M
\left[\begin{array}{c}a \\ a^\#\end{array}\right]]\nonumber \\
&&+ \frac{1}{2}L^\dagger[V,L]+\frac{1}{2}[L^\dagger,V]L
+ \tau_1^2 [V,z][z^*,V]
+\kappa zz^*
 \nonumber \\
&=& \left[\begin{array}{c}a \\ a^\#\end{array}\right]^\dagger\left(\begin{array}{c}
F^\dagger P + P F\\ 
+4 \tau_1^2 PJ\Sigma \tilde E^T \tilde E^\# \Sigma JP \\
+ \kappa \Sigma \tilde E^T \tilde E^\# \Sigma\\
\end{array}\right)\left[\begin{array}{c}a \\
a^\#\end{array}\right]\nonumber \\
&&+\tr\left(PJN^\dagger\left[\begin{array}{cc}I & 0 \\ 0 & 0 \end{array}\right]NJ\right)
\end{eqnarray}
where $F = -\imath JM-\frac{1}{2}JN^\dagger J N$. 

We now observe that applying the Schur complement to the LMI (\ref{LMI}) implies  that the matrix inequality 
\begin{equation}
\label{QMI2}
F^\dagger P + P F 
+4 \tau_1^2PJ\Sigma \tilde E^T \tilde E^\# \Sigma JP 
+ \kappa\Sigma \tilde E^T \tilde E^\# \Sigma
 < 0.
\end{equation}
will have a solution $P > 0$ of the form (\ref{Pform}).  This matrix $P$ defines a corresponding operator $V \in \mathcal{P}$ as in (\ref{quadV}). From this, it follows using (\ref{lyap_ineq3}) that 
\begin{eqnarray}
\label{dissip1a}
&&-\imath[V,\frac{1}{2}\left[\begin{array}{cc}a^\dagger &
      a^T\end{array}\right]M
\left[\begin{array}{c}a \\ a^\#\end{array}\right]]\nonumber \\
&&+ \frac{1}{2}L^\dagger[V,L]+\frac{1}{2}[L^\dagger,V]L
+ \tau_1^2[V,z][z^*,V]
\nonumber \\&&
+ \kappa zz^*
\leq \tilde \lambda
\nonumber \\
\end{eqnarray}
with 
\[
\tilde \lambda = \tr\left(PJN^\dagger\left[\begin{array}{cc}I & 0 \\ 0 & 0 \end{array}\right]NJ\right) \geq 0.
\]
Also, it follows from Lemma \ref{L2} that
\begin{eqnarray}
\label{ineq1a}
\lefteqn{-\imath[V,H] + \frac{1}{2}L^\dagger[V,L]+\frac{1}{2}[L^\dagger,V]L+W(z,z^*)}\nonumber \\
 &=& -\imath[V,f(z,z^*)]-\imath[V,\frac{1}{2}\left[\begin{array}{cc}a^\dagger &
      a^T\end{array}\right]M
\left[\begin{array}{c}a \\ a^\#\end{array}\right]]\nonumber \\
&&+ \frac{1}{2}L^\dagger[V,L]+\frac{1}{2}[L^\dagger,V]L+W(z,z^*)\nonumber \\
&=&-\imath[V,\frac{1}{2}\left[\begin{array}{cc}a^\dagger &
      a^T\end{array}\right]M
\left[\begin{array}{c}a \\ a^\#\end{array}\right]]\nonumber \\
&&+ \frac{1}{2}L^\dagger[V,L]+\frac{1}{2}[L^\dagger,V]L+W(z,z^*)\nonumber \\
&&-\imath[V,z]w_{1}^*+\imath w_{1}[z^*,V]\nonumber \\
&&-\frac{1}{2}\imath\mu w_{2}^*+\frac{1}{2}\imath w_{2}\mu^*. 
\end{eqnarray}
Furthermore, $[V,z]^* = z^*V-Vz^*=[z^*,V]$ since $V$ is self-adjoint. Therefore, for $\tau_1 > 0$
\begin{eqnarray*}
0 &\leq& \left(\tau_1[V,z]- \frac{1}{\tau_1} \imath w_{1}\right)
\left(\tau_1[V,z]- \frac{1}{\tau_1}\imath w_{1}\right)^*\nonumber \\
&=&  \tau_1^2[V,z][z^*,V]+\imath[V,z]w_{1}^*\nonumber \\
&&-\imath  w_{1}[z^*,V]+ \frac{1}{\tau_1^2}w_{1} w_{1}^*
\end{eqnarray*}
and hence
\begin{eqnarray}
\label{ineq3a}
 \lefteqn{-\imath[V,z]w_{1}^*+\imath w_{1}[z^*,V]}\nonumber \\
&\leq& \tau_1^2[V,z][z^*,V]+\frac{1}{\tau_1^2}w_{1} w_{1}^*.%\nonumber \\
\end{eqnarray}

Also, for $\tau_2 > 0$
\begin{eqnarray*}
0 &\leq& \left(\frac{\tau_2}{2}\mu- \frac{1}{\tau_2}\imath w_{2}\right)
\left(\frac{\tau_2}{2}\mu_i-\frac{1}{\tau_2} \imath w_{2i}\right)^*\nonumber \\
&=&  \frac{\tau_2^2}{4}  \mu \mu^*-\frac{\imath}{2}  w_{2}\mu^*+\frac{\imath}{2} \mu w_{2}^*
\nonumber \\
&&+  \frac{1}{\tau_2^2}w_{2}  w_{2}^*
\end{eqnarray*}
and hence
\begin{eqnarray}
\label{ineq3b}
\lefteqn{\frac{\imath}{2} w_{2}\mu^*-\frac{\imath}{2} \mu w_{2}^*}\nonumber \\
 &\leq& \frac{\tau_2^2}{4} \mu\mu^*
%\nonumber \\&&
+ \frac{1}{\tau_2^2}w_{2} w_{2}^*.%\nonumber \\
\end{eqnarray}
Also, it follows from  (\ref{sector4b}) that
\begin{equation}
\label{sector2b}
 w_{2} w_{2}^* \leq \delta_3.
\end{equation}
If we let $\tau_2^2 = \frac{2\sqrt{\delta_3}}{|\mu|}$, it follows from (\ref{ineq3b}) and (\ref{sector2b}) that
\begin{equation}
\label{ineq3c}
\frac{\imath}{2} w_{2}\mu^*-\frac{\imath}{2} \mu w_{2}^*  \leq \frac{1}{2}\sqrt{\delta_3}|\mu|+\frac{1}{2}\sqrt{\delta_3}|\mu|
 = \sqrt{\delta_3}|\mu|.
\end{equation}

Furthermore, it follows from (\ref{sector4a}) and (\ref{sector4c})  that
\begin{equation}
\label{sector2a}
 W(z,z^*)+w_{1} w_{1}^* \leq \frac{1}{\gamma_1^2} z z^* + \delta_1
\end{equation}
and 
\begin{equation}
\label{sector2c}
 w_{1} w_{1}^* \leq \frac{1}{\gamma_2^2} z z^* + \delta_2. 
\end{equation}
Combining these equations with (\ref{Wbound}), it follows that
\begin{align}
\label{sector2d}
&W(z,z^*)+\frac{1}{\tau_1^2}w_{1} w_{1}^* \nonumber\\
& \phantom{W} \leq \left\{\begin{array}{ll}
\frac{1}{\gamma_1^2} z z^* + \delta_1 & \\
+ \left(\frac{1}{\tau_1^2}-1\right)\left(\frac{1}{\gamma_2^2} z z^* 
+ \delta_2\right) & \mbox{ for } \tau_1^2 \leq 1; \\
\ \\
\frac{1}{\tau_1^2}\left(\frac{1}{\gamma_1^2} z z^* + \delta_1 \right) &\\
+ \left(1-\frac{1}{\tau_1^2}\right)\left(\frac{1}{\gamma_0^2}z z^* + \delta_0 \right)&  \mbox{ for } \tau_1^2 > 1.
\end{array}\right. \nonumber \\
\end{align}

Substituting (\ref{ineq3a}), (\ref{ineq3c}), and (\ref{sector2a})  into (\ref{ineq1a}), it follows that
\begin{eqnarray}
\label{ineq2a}
\lefteqn{-\imath[V,H] + \frac{1}{2}L^\dagger[V,L]+\frac{1}{2}[L^\dagger,V]L+W(z,z^*)}\nonumber \\
 &\leq &  -\imath[V,\frac{1}{2}\left[\begin{array}{cc}a^\dagger &
      a^T\end{array}\right]M
\left[\begin{array}{c}a \\ a^\#\end{array}\right]]\nonumber \\
&& + \frac{1}{2}L^\dagger[V,L]+\frac{1}{2}[L^\dagger,V]L\nonumber \\
&&+   \tau_1^2[V,z][z^*,V]
\nonumber \\&&
+W(z,z^*)+\frac{1}{\tau_1^2}w_{1} w_{1}^*
+\sqrt{\delta_3}|\mu|.
\end{eqnarray}
Hence, if $\tau_1^2 \leq 1$, it follows from (\ref{sector2d}) that 
\begin{align}
\label{ineq2b}
&-\imath[V,H] + \frac{1}{2}L^\dagger[V,L]+\frac{1}{2}[L^\dagger,V]L+W(z,z^*) \nonumber \\
&\phantom{-}\leq -\imath[V,\frac{1}{2}\left[\begin{array}{cc}a^\dagger &
      a^T\end{array}\right]M
\left[\begin{array}{c}a \\ a^\#\end{array}\right]]\nonumber \\
&\phantom{-=} + \frac{1}{2}L^\dagger[V,L]+\frac{1}{2}[L^\dagger,V]L+   \tau_1^2[V,z][z^*,V]
\nonumber \\
&\phantom{-=}+\left(\frac{1}{\gamma_1^2} +\left(\frac{1}{\tau_1^2}-1\right)\right) z z^* \nonumber \\
&\phantom{-=}+ \delta_1  + \left(\frac{1}{\tau_1^2}-1\right)\delta_2+\sqrt{\delta_3}|\mu|. 
\end{align}
Similarly, if $\tau_1^2 >  1$, it follows from (\ref{sector2d}) that 
\begin{align}
\label{ineq2b}
&-\imath[V,H] + \frac{1}{2}L^\dagger[V,L]+\frac{1}{2}[L^\dagger,V]L+W(z,z^*) \nonumber \\
&\phantom{-}\leq -\imath[V,\frac{1}{2}\left[\begin{array}{cc}a^\dagger &
      a^T\end{array}\right]M
\left[\begin{array}{c}a \\ a^\#\end{array}\right]]\nonumber \\
&\phantom{-=} + \frac{1}{2}L^\dagger[V,L]+\frac{1}{2}[L^\dagger,V]L+   \tau_1^2[V,z][z^*,V]
\nonumber \\
&\phantom{-=}+\left(\frac{1}{\tau_1^2\gamma_1^2} +\frac{1}{\gamma_0^2}\left(1-\frac{1}{\tau_1^2}\right)\right) z z^* \nonumber \\
&\phantom{-=}+ \frac{1}{\tau_1^2}\delta_1  + \left(1-\frac{1}{\tau_1^2}\right)\delta_0+\sqrt{\delta_3}|\mu|. 
\end{align}
Hence,
\begin{align}
\label{ineq2b}
&-\imath[V,H] + \frac{1}{2}L^\dagger[V,L]+\frac{1}{2}[L^\dagger,V]L+W(z,z^*) \nonumber \\
&\phantom{-}\leq -\imath[V,\frac{1}{2}\left[\begin{array}{cc}a^\dagger &
      a^T\end{array}\right]M
\left[\begin{array}{c}a \\ a^\#\end{array}\right]]\nonumber \\
&\phantom{-=} + \frac{1}{2}L^\dagger[V,L]+\frac{1}{2}[L^\dagger,V]L+   \tau_1^2[V,z][z^*,V]
\nonumber \\
&\phantom{-=}+\kappa z z^* \nonumber \\
&\phantom{-=}+\zeta+\sqrt{\delta_3}|\mu|
\end{align}
where $\kappa > 0$ is defined in (\ref{kappa}) and $\zeta > 0$ is defined in (\ref{zeta}). 
Then it follows from (\ref{dissip1a}) that 
\begin{align*}
&-\imath[V,H] + \frac{1}{2}L^\dagger[V,L]+\frac{1}{2}[L^\dagger,V]L + W(z,z^*)\nonumber \\
&\phantom{-} \leq \tilde \lambda + \zeta + \sqrt{\delta_3}|\mu|.
\end{align*}
From this,  it follows from Lemma \ref{L0} with $\lambda = \tilde \lambda+ \zeta + \sqrt{\delta_3}|\mu|$ that the bound 
(\ref{Cbound}) is satisfied. 
\hfill $\Box$

Note that the problem of minimizing the bound on the right hand side of (\ref{Cbound}) subject to the constraint (\ref{LMI}) can be converted into a standard LMI optimization problem which can be solved using standard LMI software; e.g., see \cite{BEFB94,BCPS:05}. 

\section{Illustrative Example}
\label{sec:example}
\begin{figure}[htbp]
\begin{center}
\includegraphics[width=8cm]{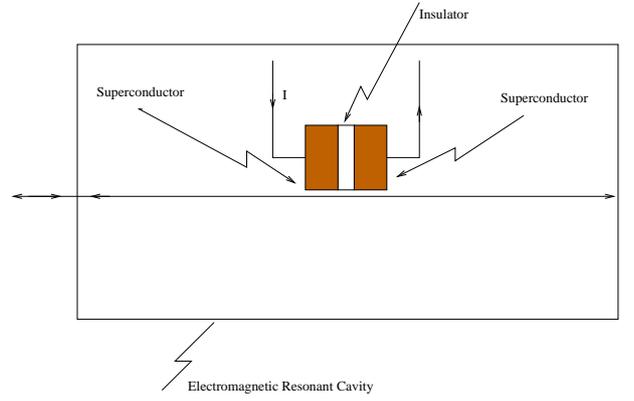}
\end{center}
\caption{Schematic diagram of a Josephson junction in a resonant cavity.}
\label{F1}
\end{figure}

To illustrate the main result of this paper, we consider an illustrative example  consisting of a Josephson junction
in an electromagnetic resonant cavity. This system was considered in the
paper \cite{PET12Aa} using a model derived from a model presented in  \cite{AS01}. The system  is illustrated in Figure \ref{F1}. 

\begin{figure}[htbp]
\begin{center}
\includegraphics[width=8cm]{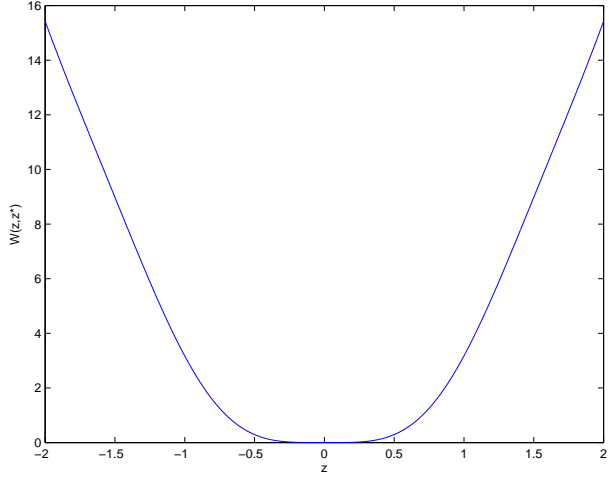}
\end{center}
\caption{Plot of non-quadratic cost  $W(z,z^*) = 4zz^* - \sin^2(z+z^*).$}
\label{F1A}
\end{figure}

In the paper \cite{PET12Aa}, a model for this system of the form considered in Section \ref{sec:systems} is derived and we consider the same model but with simplified parameter values for the purposes of this illustration. That is, we consider a Hamiltonian of the form (\ref{H1}) where
\[
M = \left[\begin{array}{cccc}1 &0& 0& 0\\
0 & 1 & -0.5 & 0 \\
0 & -0.5 & 1 & 0 \\
0 & 0 & 0 & 0 
\end{array}\right]
\]
and
\[
f(z,z^*) = - \cos(z+z^*)
\]
where $z = \frac{a_2}{\sqrt{2}}$. Hence, 
\[
\tilde E = \left[\begin{array}{cccc}0 &  \frac{1}{\sqrt{2}} &0  & 0 \end{array}\right]. 
\]
Also, we consider a coupling operator vector $L$ of the form (\ref{L})
\[
L = \left[\begin{array}{c}
4a_1\\ 4a_2
\end{array}\right].
\]
 In addition, we consider 
a non-quadratic cost function of the form (\ref{W}) where
\[
W(z,z^*) = 4zz^* - \sin^2(z+z^*) \leq 4zz^*. 
\]
Hence, we can set $\gamma_0 = \frac{1}{2}$ and $\delta_0 = 0$ in (\ref{Wbound}). A plot of the function $W(z,z^*)$ versus $z$ for a real scalar $z$  is shown in Figure \ref{F1A}. 
Furthermore, we calculate
\begin{eqnarray*}
\frac{\partial f(z,z^*)}{\partial z} &=& \sin(z+z^*)\\
\frac{\partial^2 f(z,z^*)}{\partial z^2}  &=& \cos(z+z^*).
\end{eqnarray*}
From this it follows that 
\begin{align*}
&W(z,z^*)+\frac{\partial f(z,z^*)}{\partial z}^*\frac{\partial f(z,z^*)}{\partial z} \\
&\phantom{W} = 4zz^*,
\end{align*}
and hence, (\ref{sector4a}) is satisfied with $\gamma_1 = \frac{1}{2}$ and $\delta_1 = 0$. 
Also, 
\[
\frac{\partial f(z,z^*)}{\partial z}^*\frac{\partial f(z,z^*)}{\partial z} = \sin^2(z+z^*) \leq 4 z z^*, 
\]
and hence, (\ref{sector4c}) is satisfied with $\gamma_2 = \frac{1}{2}$ and $\delta_2 = 0$. Moreover, 
\[
\quad \frac{\partial^2 f(z,z^*)}{\partial z^2}^*\frac{\partial^2 f(z,z^*)}{\partial z^2} = \cos^2(z+z^*)\leq 1, 
\]
and hence
(\ref{sector4b}) is satisfied with $\delta_3 = 1$.

We now apply Theorem \ref{T1} to find a bound on the cost (\ref{W}). This is achieved by solving the corresponding LMI optimization problem. In this case a solution to the LMI problem is found with 
\[
P = \left[\begin{array}{cccc}
   0.012      &            0       &           0      &      -0.0006\\
        0      &       0.75&            -0.0006 &        0          \\
        0     &       -0.0006&   0.012&              0      \\    
  -0.0006&       0          &        0       &      0.75       
\end{array}\right]
\]
and $\tau_1 = 0.8165$. This leads to a cost bound (\ref{Cbound}) of $\mathcal{C} \leq 6.0965$.

%\bibliography{/home/irp/Bibliog/irpnew}
%\bibliography{/home/s8504138/Bibliog/irpnew}  
%\bibliographystyle{IEEEtran}

\end{document}